\documentclass[conference]{IEEEtran}
\usepackage{amsmath,amssymb} \interdisplaylinepenalty=2500
\usepackage{color}
\usepackage{subfigure}
\usepackage{epsfig}
\usepackage{graphicx}
\usepackage{dsfont}
\usepackage{verbatim}
\usepackage{amsthm}
\usepackage{setspace}
\usepackage{mathtools}
\usepackage{enumerate}

\usepackage{tabulary}
\usepackage{booktabs}
\usepackage{bbm}

\usepackage[table]{xcolor}
\usepackage{color, colortbl}
\usepackage{multirow,bigdelim}
\usepackage{amstext}
\usepackage{array}
\usepackage{cite}

\newcount\rowc

\makeatletter
\def\ttabular{%
\hbox\bgroup
\let\\\cr
\def\rulea{\ifnum\rowc=\@ne \hrule height 1.3pt \fi}
\def\ruleb{
\ifnum\rowc=1\hrule height 1.3pt \else
\ifnum\rowc=6\hrule height \heavyrulewidth 
   \else \hrule height \lightrulewidth\fi\fi}
\valign\bgroup
\global\rowc\@ne
\rulea
\hbox to 10em{\strut \hfill##\hfill}%
\ruleb
&&%
\global\advance\rowc\@ne
\hbox to 10em{\strut\hfill##\hfill}%
\ruleb
\cr}
\def\endttabular{%
\crcr\egroup\egroup}

\newtheorem{definition}{Definition}
\newtheorem{theorem}{Theorem}
\newtheorem{lemma}[theorem]{Lemma}

\newenvironment{remark}{\textit{Remark: }}{}

\newif\ifcomments
\commentstrue 

\DeclarePairedDelimiterX{\norm}[1]{\lVert}{\rVert}{#1}

\newcommand{\Mod}[1]{\ (\text{mod}\ #1)}

\newcommand{\nFile}{K}
\newcommand{\sFile}{L}
\newcommand{\nNode}{N}
\newcommand{\nSysNode}{M}
\newcommand{\queries}{\mathcal{Q}}

\newcommand{\answer}{A}
\newcommand{\secrecy}{S}
\newcommand{\compkprime}{\bar{k'}}
\newcommand{\compk}{\bar{k}}
\newcommand{\comptheta}{\bar{\theta}}

\newcommand{\Fq}{\mathds{F}_{q}}

\newcolumntype{L}{>{$}l<{$}}

\definecolor{Gray}{gray}{0.9}
\definecolor{LightCyan}{rgb}{0.88,1,1}
\definecolor{HoneydewTwo}{rgb}{0.96,0.92,0.88}

\begin{document}

\title{Symmetric Private Information Retrieval For MDS Coded Distributed Storage}
\author{\IEEEauthorblockN{Qiwen~Wang, and~Mikael~Skoglund}
    \IEEEauthorblockA{School of Electrical Engineering, KTH Royal Institute of Technology} \vspace{-1.2em}
    \\Email: \{qiwenw, skoglund\}@kth.se \vspace{-1em}}

\maketitle
\begin{abstract}
A user wants to retrieve a file from a database without revealing the identity of the file retrieved at the database, which is known as the problem of \emph{private information retrieval (PIR)}. If it is further required that the user obtains no information about the database other than the desired file, the concept of \emph{symmetric private information retrieval (SPIR)} is introduced to guarantee privacy for both parties. In this paper, the problem of SPIR is studied for a database stored among $\nNode$ nodes in a distributed way, by using an $(N,M)$-MDS storage code. The information-theoretic capacity of SPIR, defined as the maximum number of symbols of the desired file retrieved per downloaded symbol, for the coded database is derived. It is shown that the SPIR capacity for coded database is $1-\frac{M}{N}$, when the amount of the shared common randomness of distributed nodes (unavailable at the user) is at least $\frac{M}{N-M}$ times the file size. Otherwise, the SPIR capacity for the coded database equals zero.
\end{abstract}

\section{Introduction}
Considering the scenario that a user wants to retrieve a file from a public database stored at a server, the identity of the file might be privacy-sensitive. In order to protect the identity of requested files, private information retrieval (PIR) is studied at first in \cite{chor1998private} to guarantee user privacy. To further protect the privacy of the database, symmetric private information retrieval (SPIR) is introduced~\cite{gertner1998protecting}, such that in the process of data retrieval the user obtains no more information regarding the database other than the requested file.
Inspired by \cite{chor1998private,gertner1998protecting}, the problem of PIR has been widely studied in the theoretical computer science literature, surveyed in \cite{gasarch2004survey}. In those works, the problem is studied by considering a file as a single bit and the database as a bit string. The retrieval process includes a querying phase when the user sends queries to the nodes, and a downloading phase when the nodes generate answers after receiving the queries and send back to the user. The objective is to minimize the total communication cost during both the querying phase and the downloading phase.

Recently, a series of works studies the information-theoretic limits of the communication cost of PIR problems~\cite{sun2016capacity,sun2016colluding,sun2016SPIR}. These works focus on the scenario when the file size is significantly large, and the target is to minimize the communication cost of only the downloading phase. The metric of the downloading cost is defined as the number of bits downloaded per bit of the retrieved file, and the reciprocal of which is named the PIR capacity~\cite{sun2016capacity}. The PIR capacity for a replicated database is derived in~\cite{sun2016capacity}, in which each of the $N$ (non-colluding) nodes stores a copy of the whole database. Its subsequent work~\cite{sun2016colluding} derives the PIR capacity with colluding nodes, in which case any $T$ out of $N$ nodes may collude to guess the identity of the requested file. Being the work most related to our study, another subsequent paper~\cite{sun2016SPIR} derives the capacity of SPIR in the case of a replicated database.

Considering the aspect of cost in storage systems, replicating the database results in low repair cost for node damage, but with the expense of high storage cost. Coded storage is proposed to utilize the tradeoff between storage cost and repair cost~\cite{dimakis2010network}. By using erasure codes, each node only stores a fraction of the whole database, hence reducing the storage cost.
The first work on PIR for coded database known to us appears in~\cite{shah2014one}. In~\cite{shah2014one}, the authors show that by downloading one extra bit besides the amount of the file size, user privacy can be guaranteed. However, to achieve this low downloading cost, the number of storage nodes needs to grow with the file size, which can be impractical in some storage systems.
Later,~\cite{fazeli2015pir} also considers PIR with coded storage, and focuses on reducing the storage overhead.
In~\cite{chan2015private}, PIR for coded databases is investigated, and the tradeoff between storage cost and downloading cost is analyzed. Subsequently in~\cite{tajeddine2016private}, explicite storage and communication schemes to achieve PIR with MDS storage codes are presented, matching the tradeoff derived in~\cite{chan2015private}.
It is worth noting that in the recent work of~\cite{banawan2016capacity}, the capacity of PIR for coded database is settled, which improves the results in~\cite{chan2015private} and~\cite{tajeddine2016private}.

In this work, the problem of SPIR is studied for coded databases, where the database is stored at the nodes by an MDS storage code. 
We show that in order to guarantee SPIR in the non-trivial context, {\it e.g.,} the number of files in the database is greater than or equal to two, nodes need to share common randomness which is independent to the database and meanwhile unavailable to the user. This result is in analogy with that in~\cite{sun2016SPIR} for the uncoded database.
In particular, we derive a lower bound on the amount of common randomness needed to assure positive SPIR retrieval rate. 
Furthermore, the capacity of SPIR for the $(N,M)$-MDS coded database is found.
We note that the replicated database is a special case of the coded database with $(N,1)$-MDS code.
Therefore, our result includes that in~\cite{sun2016SPIR} for the replicated database as a special case with $M=1$.

%
%
%
%

\section{Model}
\subsection{Notations}
Let $[1:N]$ denote the set $\{1,2, \dots, N\}$ and $[M,N]$ denote $\{M, M+1, \dots, N\}$ for $M \leq N$. For the sake of brevity, denote the set of random variables $\{ X_1, X_2, \dots, X_N\}$ by $X_{[1:N]}$ . Let $e_i$ denote the unit vector with a one at the $i$th entry, and zeros at all other entries, the length of which is not specified when there is no ambiguity.

\subsection{Problem Description}
\noindent{\bf Database:}
A database comprises $\nFile$ independent files, denoted by $W_1, \dots, W_{\nFile}$. Each file consists of $\sFile$ symbols drawn independently and uniformly from the finite field $\Fq$. Therefore, for any $k \in [1:\nFile]$, 
\begin{equation}
H(W_k)=L \log{q} \quad ; \quad H(W_1, \dots, W_{\nFile}) = KL \log{q}. \nonumber
\end{equation}

\noindent{\bf Storage:} 
The database is stored in a distributed storage system consisting of $\nNode$ nodes by an $(\nNode, \nSysNode)$-MDS storage code.  The data stored at the $\nNode$ nodes are denoted by $D_1, \dots, D_{\nNode}$. Note that with an $(\nNode, \nSysNode)$-MDS storage code, for any $\nSysNode$ nodes $\{n_1, \dots, n_M\} \in [1:N]$, the data they store $D_{n_1},\dots, D_{n_M}$ are linearly and stochastically independent.
Furthermore, every $\nSysNode$ nodes can exactly recover the whole database, {\it i.e.,}
\begin{equation}
H(D_{n_1},\dots, D_{n_M}) = H(W_1, \dots, W_{\nFile}) = KL \log{q}, \nonumber
\end{equation}
\begin{equation}
H(W_1,\ldots,W_K|D_{n_1},\ldots,D_{n_M}) = 0. \nonumber
\end{equation}

\noindent{\bf User queries:}
A user wants to retrieve a file $W_{\theta}$ with index $\theta$ from the database, $\theta \in [1:\nFile]$. 
The MDS storage code is known to the user. In addition to this, the user has no knowledge of the stored data.
Based on the desired file index $\theta$, the user sends queries to all nodes, where the query received by node $n$ 
is denoted by $Q_{n}^{\theta}$. Let $\queries = [Q_{n}^{\theta}]_{n \in [1:\nNode], \theta \in [1:\nFile]}$ denote the complete query scheme, namely, the collection of all queries under all cases of desired file index.

\noindent{\bf Node answers:}
Based on the received query $Q_n^{\theta}$, the stored data $D_n$, and some common randomness $\secrecy$ shared among all nodes, each node sends an answer $\answer_{n}^{\theta}$ to the user.
The common randomness is utilized to protect database-privacy~\eqref{eqn:data_privacy} below.

\noindent{\bf SPIR:}
With the received answers $\answer_{[1:\nNode]}^{\theta} = \{ \answer_{1}^{\theta}, \dots, \answer_{\nNode}^{\theta} \}$ and based on the complete query scheme $\queries$, the user shall be able to decode the requested file $W_{\theta}$ with zero error. The nodes do not communicate, that is, they share no information regarding their stored data and the queries they receive. The only information shared among the nodes is some common randomness, denoted by $\secrecy$, which is independent of the database and unavailable to the user.
Two privacy constraints must be satisfied for SPIR:
\begin{itemize}
\item \emph{User-privacy:} the nodes shall not be able to obtain any information regarding the identity of the requested file, {\it i.e.,}
	\begin{equation}
		I(\theta ; Q_n^{\theta}, A_n^{\theta}, D_n, S ) = 0, \quad \forall n \in [1:\nNode]. \label{eqn:user_privacy}
	\end{equation}
\item \emph{Database-privacy:} the user shall learn no information regarding other files in the database, that is, defining $W_{\comptheta}= \{ W_1, \dots, W_{\theta-1}, W_{\theta+1}, \dots, W_{\nFile}\}$, 
	\begin{equation}
		I(W_{\comptheta} ; \answer_{[1:\nNode]}^{\theta}, \queries, \theta) = 0. \label{eqn:data_privacy}
	\end{equation}
\end{itemize}

We use the same definition of SPIR rate and capacity as in~\cite{sun2016SPIR} for characterization of the performance of SPIR schemes.
\begin{definition}[SPIR Rate and Capacity]
The SPIR rate is the number of information bits of the requested file retrieved per downloaded answer bit, that is,
\begin{equation}
R_{\mathrm{SPIR}}^{(N,M)\mathrm{-MDS}} \triangleq \frac{H(W_{\theta})}{\sum_{n=1}^{\nNode} H(A_n^{\theta})}. \nonumber
\end{equation}
The capacity $C_{\mathrm{SPIR}}^{(N,M)\mathrm{-MDS}}$ is the supremum of $R_{\mathrm{SPIR}}^{(N,M)\mathrm{-MDS}} $ over all SPIR schemes for $(N,M)$-MDS storage codes.
\end{definition}

\begin{definition}[Secrecy Rate]
The secrecy rate is the amount of common randomness shared by the storage nodes relative to the file size, that is
\begin{equation}
\rho_{\mathrm{SPIR}}^{(N,M)\mathrm{-MDS}} \triangleq \frac{H(S)}{H(W_{\theta})}. \nonumber
\end{equation}
\end{definition}

\section{Main Result} \label{sec:main}
When there is only one file in the database, {\it i.e.} $K=1$, database-privacy is guaranteed automatically, because there is no other file to protect from the user in the database. Therefore, the SPIR problem reduces to PIR problem, and from~\cite{banawan2016capacity}, the capacity is $1$ regardless of the rate of the MDS-storage code. When $K \geq 2$, SPIR is non-trivial and our main result is summarized below.

\begin{theorem}
For symmetric private information retrieval from a database with $K \geq 2$ files which are stored at $N$ nodes with an $(N,M)$-MDS storage code, the capacity is 
\begin{equation}
C_{\mathrm{SPIR}}^{(N,M)\mathrm{-MDS}} = 
\begin{cases}
1-\frac{M}{N}, & \text{if } \rho_{\mathrm{SPIR}}^{(N,M)\mathrm{-MDS}} \geq \frac{M}{N-M}\\
0, & \text{otherwise}
\end{cases}
. \nonumber
\end{equation} 
\label{thm:main}
\end{theorem} 

\begin{remark}
When $M=1$, that is, every node stores the whole database, our result reduces to Theorem 1 in~\cite{sun2016SPIR} for replicated databases. In~\cite{banawan2016capacity}, the authors show that the PIR capacity with MDS storage codes is $(1+\frac{M}{N}+\cdots+\frac{M^{K-1}}{N^{K-1}})^{-1}$.
(We translate their result into our notation.) It can be observed that as the number of files $K$ tends to infinity, 
their PIR capacity approaches our SPIR capacity.
The intuition is that, when the number of files increases, the penalty in the downloading rate to protect database-privacy decays.
When there are asymptotically infinitely many files, the information rate the user can learn about the database from finite downloaded symbols vanishes.
\end{remark}

\section{Converse} \label{sec:converse}
In this section, we show the converse part of Theorem~\ref{thm:main}. That is, to achieve SPIR with an $(\nNode,\nSysNode)$-MDS storage code, the nodes need to share at least $\frac{\nSysNode}{\nNode-\nSysNode} L$ secrecy symbols (Theorem~\ref{thm:secrecy}), and the user needs to download at least $\frac{\nNode}{\nNode-\nSysNode} L$ symbols (Theorem~\ref{thm:converse}). Recall that $L$ is the file size. We first show Lemmas~\ref{thm:Lemma2}-\ref{thm:eqn35} below, which will be used in the proofs of Theorems~\ref{thm:converse} and~\ref{thm:secrecy}.
\begin{lemma} \label{thm:Lemma2}
For any $\nSysNode$ nodes $n_{[1:M]} \triangleq \{ n_1, \dots, n_{\nSysNode} \} \subset [1:\nNode]$, 
	\begin{equation}
		H(A_{n_{[1:M]}}^{k} | \queries, W_k, Q_{n_{[1:M]}}^{k}) = H(A_{n_{[1:M]}}^{k} | W_k, Q_{n_{[1:M]}}^{k}). \nonumber
	\end{equation}	
\end{lemma}
\noindent {\it Proof:}
We first show that $H(A_{n_{[1:M]}}^{k} | W_k, Q_{n_{[1:M]}}^{k}) \leq  H(A_{n_{[1:M]}}^{k} | \queries, W_k, Q_{n_{[1:M]}}^{k})$, as follows
\begin{align*}
& H(A_{n_{[1:M]}}^{k} | W_k, Q_{n_{[1:M]}}^{k}) - H(A_{n_{[1:M]}}^{k} | \queries, W_k, Q_{n_{[1:M]}}^{k}) \\
& = I(A_{n_{[1:M]}}^{k}; \queries | W_k, Q_{n_{[1:M]}}^{k}) \\
& \leq I(A_{n_{[1:M]}}^{k}, W_1,\dots, W_{\nFile}, \secrecy ; \queries | W_k, Q_{n_{[1:M]}}^{k}) \\
& = I(W_1,\dots, W_{\nFile}, \secrecy ; \queries | W_k, Q_{n_{[1:M]}}^{k}) + \\
& \qquad I(A_{n_{[1:M]}}^{k} ; \queries |  W_1,\dots, W_{\nFile}, \secrecy ,W_k, Q_{n_{[1:M]}}^{k}) \\
& \stackrel{(a)}{=} I(W_1,\dots, W_{\nFile}, \secrecy ; \queries | W_k, Q_{n_{[1:M]}}^{k}) \\
& \leq I(W_1,\dots, W_{\nFile}, \secrecy ; \queries) = 0,
\end{align*}
where equality $(a)$ holds because the answers are deterministic functions of the database, the common randomness, and the queries.
In the last step, $I(W_1,\dots, W_{\nFile}, \secrecy ; \queries) = 0$ holds because the queries do not depend on the database and the common randomness.

On the other hand, it is immediate that $H(A_{n_{[1:M]}}^{k} | W_k, Q_{n_{[1:M]}}^{k}) \geq  H(A_{n_{[1:M]}}^{k} | \queries, W_k, Q_{n_{[1:M]}}^{k})$. Therefore, $H(A_{n_{[1:M]}}^{k} | W_k, Q_{n_{[1:M]}}^{k}) = H(A_{n_{[1:M]}}^{k} | \queries, W_k, Q_{n_{[1:M]}}^{k})$.
\hfill $\Box$

\begin{lemma} \label{thm:Lemma1}
For any $\nSysNode$ nodes $n_{[1:M]} \triangleq \{ n_1, \dots, n_{\nSysNode} \} \subset [1:\nNode]$, 
	\begin{equation} 
		H(A_{n_{[1:M]}}^{k} |  Q_{n_{[1:M]}}^{k}) = H(A_{n_{[1:M]}}^{k'}|  Q_{n_{[1:M]}}^{k'}), \label{eqn:Lemma1_2}
	\end{equation}
	\begin{equation} 
		H(A_{n_{[1:M]}}^{k} | W_k, Q_{n_{[1:M]}}^{k}) = H(A_{n_{[1:M]}}^{k'} | W_k, Q_{n_{[1:M]}}^{k'}). \label{eqn:Lemma1_1}
	\end{equation}	
\end{lemma}
\noindent{\it Proof:}
{\it Proof of~\eqref{eqn:Lemma1_2}:}

From user-privacy~\eqref{eqn:user_privacy}, $I(\theta; A_n^{\theta}, Q_n^{\theta}) = 0$, hence $H(A_n^{k}, Q_n^{k})=H(A_n^{k'}, Q_n^{k'})$. Similarly, $I(\theta;  Q_n^{\theta}) = 0$, therefore $H( Q_n^{k})=H(Q_n^{k'})$.
From the above,  we have that $H(A_n^{k}| Q_n^{k})=H(A_n^{k'}| Q_n^{k'})$.

W.o.l.g., we choose the size-$\nSysNode$ node set $n_{[1:M]}$ to be $\{1, \dots, \nSysNode\}$. 
For an $(\nNode, \nSysNode)$-MDS storage code, the data stored at any set of $M$ nodes are linearly independent. Furthermore, because the files in the database are statistically independent, the data stored at any set of $M$ nodes are also statistically independent. (See Lemma 1 in~\cite{sun2016colluding} and Lemma 2 in~\cite{banawan2016capacity} for a proof.) 
For any node $n$, the answer $A_n^k$ is a deterministic function of the query $Q_n^k$, the common randomness $\secrecy$, and the stored data $D_k$. 
Given the queries and the common randomness, the randomness of the $M$ answers only lies in the stored data of the $M$ nodes, which are statistically independent. Hence,
\begin{equation}
	H(A_{[1:\nSysNode]}^k | Q_{[1:\nSysNode]}^k, \secrecy ) = \sum_{n=1}^{\nSysNode} H(A_n^k | Q_n^k, \secrecy ). \nonumber
\end{equation}

Because the user shall not obtain any information of the common randomness $S$ from the queries and answers, $S$ should be independent of the queries and answers. Therefore, $H(A_{[1:\nSysNode]}^k | Q_{[1:\nSysNode]}^k, \secrecy ) = H(A_{[1:\nSysNode]}^k | Q_{[1:\nSysNode]}^k) + H(\secrecy | A_{[1:\nSysNode]}^k , Q_{[1:\nSysNode]}^k) - H(\secrecy | Q_{[1:\nSysNode]}^k) = H(A_{[1:\nSysNode]}^k | Q_{[1:\nSysNode]}^k)$.
Similarly, we have that $H(A_n^k | Q_n^k, \secrecy ) = H(A_n^k | Q_n^k ) + H(\secrecy | A_n^k , Q_n^k) - H(\secrecy | Q_n^k) = H(A_n^k | Q_n^k ) $. Hence,
\begin{align*}
H(A_{[1:\nSysNode]}^k | Q_{[1:\nSysNode]}^k) 
& = H(A_{[1:\nSysNode]}^k | Q_{[1:\nSysNode]}^k, \secrecy ) \\
& = \sum_{n=1}^{\nSysNode} H(A_n^k | Q_n^k, \secrecy ) =  \sum_{n=1}^{\nSysNode}  H(A_n^k | Q_n^k ) \\
& = \sum_{n=1}^{\nSysNode} H(A_n^{k'} | Q_n^{k'} ) = H(A_{[1:\nSysNode]}^{k'} | Q_{[1:\nSysNode]}^{k'}). 
\end{align*}

{\it Proof of~\eqref{eqn:Lemma1_1}:}
Let the random variable $D_n^k$ denote the randomness of $D_n$ after fixing $W_{\compk}$, that is, the part of randomness of file $W_k$ stored at node $n$. By user-privacy~\eqref{eqn:user_privacy}, $I(\theta ; Q_n^{\theta}, A_n^{\theta}, D_n^k) = 0$, we have that $H(Q_n^{k}, A_n^{k}, D_n^k) = H(Q_n^{k'}, A_n^{k'}, D_n^k)$ and $H(Q_n^{k}, D_n^k) = H(Q_n^{k'}, D_n^k)$. Hence, $H( A_n^{k}| Q_n^{k},D_n^k) = H( A_n^{k'}|Q_n^{k'}, D_n^k)$.

The answer $A_n^k$ is a deterministic function of the query $Q_n^k$, the common randomness $\secrecy$, and the stored data $D_n$. We argue above that the data stored at any set of $M$ nodes are statistically independent. After fixing the file $W_{k}$, the data stored at $M$ nodes, which depends only on the randomness of the other $K-1$ files $W_{\compk}$, are still statistically independent. Therefore,
\begin{equation}
	H(A_{[1:\nSysNode]}^k | Q_{[1:\nSysNode]}^k, D_{[1:\nSysNode]}^k, \secrecy ) = \sum_{n=1}^{\nSysNode} H(A_n^k | Q_n^k, D_n^k, \secrecy ). \nonumber
\end{equation}
The randomness relating to file $W_k$ stored in $\nSysNode$ nodes recovers $W_k$, {\it i.e.,} $D_{[1:\nSysNode]}^k = W_k$.

Because the common randomness is independent of the queries $\queries$, answers $A_{[1:N]}^k$, and $W_k$ which the user can decode, with similar calculations as in the proof for~\eqref{eqn:Lemma1_2}, we can eliminate $S$ in the conditions. Hence, $H(A_{[1:\nSysNode]}^k | Q_{[1:\nSysNode]}^k, W_k) = \sum_{n=1}^{\nSysNode} H(A_n^k | Q_n^k, D_n^k)$.

To show that $H(A_{[1:\nSysNode]}^{k'} | Q_{[1:\nSysNode]}^{k'}, W_k) = \sum_{n=1}^{\nSysNode} H(A_n^{k'} | Q_n^{k'}, D_n^k)$, notice that because all the files are statistically independent, by fixing $W_k$, it is equivalent to reducing the database to $\nFile-1$ files. Hence, the equality holds by~\eqref{eqn:Lemma1_2}.
Therefore,
\begin{align*}
	H(A_{[1:\nSysNode]}^k | Q_{[1:\nSysNode]}^k, W_k) 
	& = \sum_{n=1}^{\nSysNode} H(A_n^k | Q_n^k, D_n^k) \\
	& = \sum_{n=1}^{\nSysNode} H(A_n^{k'} | Q_n^{k'}, D_n^k) \\
	& = H(A_{[1:\nSysNode]}^{k'} | Q_{[1:\nSysNode]}^{k'}, W_k). 
\end{align*}
\hfill $\Box$

\begin{lemma} \label{thm:eqn35}
For any $\nSysNode$ nodes $n_{[1:M]} \triangleq   \{ n_1, \dots, n_{\nSysNode} \} \subset [1:\nNode]$, 
	\begin{equation}
		H(A_{n_{[1:M]}}^{k} | W_k, Q_{n_{[1:M]}}^{k}) = H(A_{n_{[1:M]}}^{k'} |  Q_{n_{[1:M]}}^{k'}). \nonumber
	\end{equation}	
\end{lemma}
\noindent{\it Proof:}
By database-privacy~\eqref{eqn:data_privacy}, $ I(W_{\compkprime} ; A_{[1:\nNode]}^{k'}, \queries) = 0$.
For $k \neq k'$, $W_k \in W_{\compkprime}$. W.o.l.g., choose the size-$\nSysNode$ node set to be $\{1, \dots, \nSysNode\}$,
\begin{align*}
0
& = I(W_{k} ; A_{[1:\nSysNode]}^{k'}, Q_{[1:\nSysNode]}^{k'}) \\
& = I(W_{k} ; A_{[1:\nSysNode]}^{k'}| Q_{[1:\nSysNode]}^{k'}) +  I(W_{k} ; Q_{[1:\nSysNode]}^{k'}) \\
& \stackrel{(a)}{=}   I(W_{k} ; A_{[1:\nSysNode]}^{k'}| Q_{[1:\nSysNode]}^{k'}) \\
& = H(A_{[1:\nSysNode]}^{k'}| Q_{[1:\nSysNode]}^{k'}) - H(A_{[1:\nSysNode]}^{k'}| W_k, Q_{[1:\nSysNode]}^{k'}) \\
& \stackrel{(b)}{=} H(A_{[1:\nSysNode]}^{k'}| Q_{[1:\nSysNode]}^{k'}) - H(A_{[1:\nSysNode]}^{k}| W_k, Q_{[1:\nSysNode]}^{k}) ,
\end{align*}
where equality $(a)$ holds because $W_k$ is independent of the queries, and equality $(b)$ follows by~\eqref{eqn:Lemma1_1} in Lemma~\ref{thm:Lemma1}.
\hfill $\Box$

\begin{theorem} \label{thm:converse}
The SPIR rate for a database stored with an $(N,M)$-MDS storage code is bounded from above by
	\begin{equation}
		R_{\mathrm{SPIR}}^{(N,M)\mathrm{-MDS}} \leq 1-\frac{M}{N}. \nonumber
	\end{equation}
\end{theorem}
\noindent{\it Proof:} For any file $W_k$, $k \in [1:K]$,
\begin{align*}
H(W_k) 
& = H(W_k|\queries) \\
& \stackrel{(a)}{=} H(W_k|\queries) - H(W_k|A_{[1:\nNode]}^k, \queries) \\
& = I(W_k ; A_{[1:\nNode]}^k| \queries) \\
& = H(A_{[1:\nNode]}^k | \queries) - H(A_{[1:\nNode]}^k | W_k, \queries) \\
& \leq H(A_{[1:\nNode]}^k | \queries) - H(A_{n_{[1:M]}}^{k} | W_k, \queries, Q_{n_{[1:M]}}^{k}) \\
& \stackrel{(b)}{=} H(A_{[1:\nNode]}^k | \queries) - H(A_{n_{[1:M]}}^{k} | W_k, Q_{n_{[1:M]}}^{k}) \\
& \stackrel{(c)}{=} H(A_{[1:\nNode]}^k| \queries) - H(A_{n_{[1:M]}}^{k'} |  Q_{n_{[1:M]}}^{k'}) \\
& \stackrel{(d)}{=} H(A_{[1:\nNode]}^k | \queries) - H(A_{n_{[1:M]}}^{k} |  Q_{n_{[1:M]}}^{k}) \\
& \leq H(A_{[1:\nNode]}^k | \queries) - H(A_{n_{[1:M]}}^{k} | \queries) \\
& \leq  H(A_{n_{[1:M]}}^{k} | \queries) + \sum_{n \in [1:\nNode] \setminus n_{[1:\nSysNode]} } H(A_n^k | \queries) - \\
& \qquad  H(A_{n_{[1:M]}}^{k} | \queries) \\
& \leq \sum_{n \in [1:\nNode] \setminus n_{[1:\nSysNode]} } H(A_n^k) \\
& \stackrel{(e)}{=} \frac{\nNode - \nSysNode}{\nNode} \cdot \sum_{n=1}^{\nNode} H(A_n^k)
\end{align*}
Equality $(a)$ holds because from all the answers and the queries, the user should be able to decode $W_k$, hence $H(W_k|A_{[1:\nNode]}^k, \queries)=0$.
Equalities $(b)$ and $(c)$ follow from Lemma~\ref{thm:Lemma2} and Lemma~\ref{thm:eqn35}.
Equality $(d)$ follows from~\eqref{eqn:Lemma1_2} in Lemma~\ref{thm:Lemma1}.
Step $(e)$ is because $\{n_1, \dots, n_{\nSysNode}\}$ can be any size $\nSysNode$ index set from $[1:\nNode]$. Hence by symmetry, $\sum_{n \in [1:\nNode] \setminus n_{[1:\nSysNode]} } H(A_n^k) = \frac{\nNode - \nSysNode}{\nNode} \cdot \sum_{n=1}^{\nNode} H(A_n^k)$.

Therefore, $R_{\mathrm{SPIR}}^{(N,M)\mathrm{-MDS}} = \frac{H(W_k)}{\sum_{n=1}^{\nNode} H(A_n^k)} \leq 1 - \frac{M}{N}$.
\hfill $\Box$

\begin{theorem} \label{thm:secrecy}
The secrecy rate for SPIR with an $(N,M)$-MDS storage code needs to be at least 
	\begin{equation}
		\rho_{\mathrm{SPIR}}^{(N,M)\mathrm{-MDS}}  \geq \frac{\nSysNode}{\nNode-\nSysNode} . \nonumber
	\end{equation}
\end{theorem}
\noindent{\it Proof:} By database-privacy~\eqref{eqn:data_privacy},
\begin{align*}
0 
& = I(W_{\compk} ; A_{[1:\nNode]}^k, \queries) \\
& =  I(W_{\compk} ; A_{[1:\nNode]}^k | \queries) + I(W_{\compk} ; \queries) \\
& =  I(W_{\compk} ; A_{[1:\nNode]}^k | \queries) \\
& = H(W_{\compk} | \queries) - H(W_{\compk} | A_{[1:\nNode]}^k  , \queries) \\
& \stackrel{(a)}{=} H(W_{\compk} | \queries, W_k) - H(W_{\compk} | A_{[1:\nNode]}^k  , \queries, W_k) \\
& = I(W_{\compk} ; A_{[1:\nNode]}^k  | \queries, W_k) \\
& \geq I(W_{\compk} ; A_{n_{[1:M]}}^{k} | \queries, W_k) \\
& \stackrel{(b)}{=} H(A_{n_{[1:M]}}^{k} | \queries, W_k) - H(A_{n_{[1:M]}}^{k} | \queries, W_{[1:\nFile]}) + \\
& \qquad H(A_{n_{[1:M]}}^{k} | \queries, W_{[1:\nFile]}, \secrecy) \\
& = H(A_{n_{[1:M]}}^{k} | \queries, W_k) - I(\secrecy ; A_{n_{[1:M]}}^{k} | \queries, W_{[1:\nFile]}) \\
& \geq  H(A_{n_{[1:M]}}^{k} | \queries, W_k, Q_{n_{[1:M]}}^{k}) - H(\secrecy) \\
& \stackrel{(c)}{=} H(A_{n_{[1:M]}}^{k} |  W_k, Q_{n_{[1:M]}}^{k}) - H(\secrecy) \\
& \stackrel{(d)}{=} H(A_{n_{[1:M]}}^{k'} | Q_{n_{[1:M]}}^{k'}) - H(\secrecy) \\
&  \stackrel{(e)}{=} H(A_{n_{[1:M]}}^{k} | Q_{n_{[1:M]}}^{k}) - H(\secrecy) \\
& \geq H(A_{n_{[1:M]}}^{k} | \queries) - H(\secrecy)
\end{align*}
Equality $(a)$ holds because  $W_k$ is independent of other files $W_{\compk}$, and from all the answers $A_1^k, \dots, A_{\nNode}^k $ and the queries $\queries$ the user can decode $W_k$.
Equality $(b)$ holds because the answers $A_{n_1}^k, \dots, A_{n_\nSysNode}^k $ are deterministic functions of the queries $\queries$, the database $W_1, \dots, W_{\nFile}$, and the common randomness $\secrecy$.
Equalities $(c)$ and $(d)$ follow from Lemma~\ref{thm:Lemma2} and Lemma~\ref{thm:eqn35}.
Equality $(e)$ follows from~\eqref{eqn:Lemma1_2} in Lemma~\ref{thm:Lemma1}.

Because $\{n_1, \dots, n_{\nSysNode}\}$ can be any size $\nSysNode$ index set from $[1:\nNode]$, 
\begin{equation}
	{N \choose M} H(A_{n_{[1:M]}}^{k}|\queries) \geq { {N-1} \choose {M-1}} H(A_{[1:\nNode]}^k | \queries), \nonumber
\end{equation}
hence $H(A_{[1:\nNode]}^k | \queries) \leq \frac{N}{M} \cdot  H(A_{n_{[1:M]}}^{k} |\queries) $.
From Theorem~\ref{thm:converse}, $H(W_k) \leq H(A_{[1:\nNode]}^k | \queries) - H(A_{n_{[1:M]}}^{k} | \queries) $, therefore $H(W_k) \leq (\frac{N}{M} - 1) H(A_{n_{[1:M]}}^{k} | \queries) $.

Hence, $H(\secrecy) \geq H(A_{n_{[1:M]}}^{k} | \queries) \geq \frac{\nSysNode}{\nNode - \nSysNode} \cdot H(W_k)$ and $\rho_{\mathrm{SPIR}}^{(N,M)\mathrm{-MDS}} = \frac{H(S)}{H(W_k)} \geq \frac{\nSysNode}{\nNode - \nSysNode}$.
\hfill $\Box$

\section{Achievability}
In this section, we present a scheme which achieves the maximum SPIR rate and lowest secrecy rate in Section~\ref{sec:converse}. Specifically, the user is able to decode the desired file successfully and privately by downloading $\frac{N}{N-M}L$ symbols, and obtains no further information regarding the database with $\frac{M}{N-M}L$ uniformly random symbols shared among the nodes.
The achievable scheme is revised from the scheme in~\cite{tajeddine2016private} by adding common randomness. We reprise the details with our notations.
The main concepts used in the construction are, 
\begin{itemize}
\item The user hides the identity of the desired file in randomly generated queries, such that the queries appear statistically uniformly random to the nodes.
\item The nodes add shared random symbols that are independent of the database and unavailable to the user in the answers to protect the content of other files. The random symbols are added according to the storage code construction for successful decoding.
\item The user downloads the lowest possible number of symbols to construct a linear system that is solvable. The unknowns are symbols of the requested file, and some function outputs generated from queries, stored data and common randomness.
\end{itemize}

\noindent{\bf Database:}
W.o.l.g, assume each file consists of $\sFile = (\nNode - \nSysNode) \nSysNode$ symbols. Specifically, 
\begin{equation}
W_k = 
\begin{bmatrix}
w_{k,1}^1 & w_{k,2}^1 & \dots & w_{k,\nSysNode}^{1} \\
w_{k,1}^2 & w_{k,2}^2 & \dots & w_{k,\nSysNode}^{2} \\
\vdots      & \vdots     & \ddots & \vdots                                    \\
w_{k,1}^{\nNode - \nSysNode} & w_{k,2}^{\nNode - \nSysNode} & \dots & w_{k,\nSysNode}^{\nNode - \nSysNode}
\end{bmatrix}
, \nonumber
\end{equation}
where $w_{k,n}^j$ denotes the $j$th symbol in the part of file $W_k$ that is stored at node-$n$ in the systematic storage code, which is described in more detail below.

\noindent {\bf Storage:}
We use a systematic $(\nNode,\nSysNode)$-MDS storage code, as presented in Table~\ref{tb:storage}. The first $\nSysNode$ nodes are systematic nodes which store independent pieces of the files. The remaining $\nNode-\nSysNode$ nodes are parity nodes which store linear combinations of the symbols at systematic nodes. In Table~\ref{tb:storage}, for $n \in [M+1,N]$, $LC^{n}(\cdot)$ denotes the linear combination at the parity node $n$, the input of which are $w_{k,[1:M]}^j = \{w_{k,1}^j, \dots, w_{k,M}^j\}$. 
The vector stored at node $n$ is denoted by $D_n$.

\begin{table}[t!]
\hskip-1.0cm
\resizebox{10cm}{!}{
\begin{tabular}{c|c|c|c|c|c|c|c|}
\cline{2-8}
 &  node $1$       &  node $2$    & $\dots$  & node $M$     &  node $M+1$                         &  $\dots$ &      node $N$ \\
 &  $D_1$   &  $D_2$  &   $\dots$ & $D_M$ & $D_{M+1}$ & $\dots$ &   $D_{N}$ \\
\cmidrule{2-8} \morecmidrules\cmidrule{2-8}
 \ldelim\{{3}{6mm}[$W_1$] & $w_{1,1}^1$ & $w_{1,2}^1$ & $\dots $ & $w_{1,M}^1$ & $LC^{M+1}(w_{1,[1:M]}^1)$ & $\dots$ & $LC^{N}(w_{1,[1:M]}^1)$ \\
  &  $\vdots$    &    $\vdots $   & $\vdots$&   $\vdots$     &    $\vdots$                             & $\vdots$ &      $\vdots$                   \\
&  $w_{1,1}^{N-M}$ & $w_{1,2}^{N-M}$ & $\dots $ & $w_{1,M}^{N-M}$ & $LC^{M+1}(w_{1,[1:M]}^{N-M})$ & $\dots$ & $LC^{N}(w_{1,[1:M]}^{N-M})$ \\
\cline{2-8}
  &  $\vdots$    &    $\vdots $   & $\vdots$&   $\vdots$     &    $\vdots$                             & $\vdots$ &      $\vdots$                   \\
\cline{2-8}
 \ldelim\{{3}{7mm}[$W_K$] & $w_{K,1}^1$ & $w_{K,2}^1$ & $\dots $ & $w_{K,M}^1$ & $LC^{M+1}(w_{K,[1:M]}^1)$ & $\dots$ & $LC^{N}(w_{K,[1:M]}^1)$ \\
 &  $\vdots$    &    $\vdots $   & $\vdots$&   $\vdots$     &    $\vdots$                             & $\vdots$ &      $\vdots$                   \\
  & $w_{K,1}^{N-M}$ & $w_{K,2}^{N-M}$ & $\dots $ & $w_{K,M}^{N-M}$ & $LC^{M+1}(w_{K,[1:M]}^{N-M})$ & $\dots$ & $LC^{N}(w_{K,[1:M]}^{N-M})$ \\ 
 \cline{2-8}
\end{tabular}
}
\caption{Systematic $(N,M)$-MDS storage code.}
\vspace{-1cm}
\label{tb:storage}
\end{table}

\noindent {\bf Common randomness:}
All nodes share $\nSysNode^2$ uniformly random symbols from $\Fq$, denoted by 
\begin{equation}
S = 
\begin{bmatrix}
S_{1,1} & S_{1,2} & \dots & S_{1,M} \\
\vdots      & \vdots     & \ddots & \vdots                                    \\
S_{M,1} & S_{M,2} & \dots & S_{M,M}
\end{bmatrix}
, \nonumber
\end{equation}
which are independent of the database and unavailable to the user.

For the details of the queries and answers, w.o.l.g, assume the desired file is $W_1$. The user generates $\nSysNode$ uniformly random vectors $U_1, \dots, U_{\nSysNode}$ of length $(\nNode-\nSysNode)\nFile$ over $\Fq$.  The detailed achievable scheme is presented in two orthogonal cases as follows. 
\begin{itemize}
\item {\bf Case 1} ($\nNode-\nSysNode \leq \nSysNode$)

\noindent {\bf Queries:}
The query to each node consists of $\nSysNode$ vectors over $\Fq$ as shown in Table~\ref{tb:query_case1}.
Recall that $e_i$ denotes the unit vector with a one at the $i$th entry, and zeros at all other entries.
Specifically, for systematic nodes, $\nNode - \nSysNode$ out of the $\nSysNode$ query vectors retrieve the $\nNode - \nSysNode$ symbols of $W_1$ stored at each node by adding the unit vectors, in a shifted way among all systematic nodes. The queries to the parity nodes are just the $\nSysNode$ random vectors $U_1, \dots, U_{\nSysNode}$.
\begin{table*}[t]
\centering
\resizebox{13cm}{!}{
\begin{tabular}{|c|c|c|c|c|c|c|}
\hline
$Q_1^1:$  & $Q_2^1: $ & $\dots$ &$Q_{\nSysNode}^1: $ &$Q_{\nSysNode+1}^1: $ & $\dots$ & $Q_{\nNode}^1: $ \\
\hline \hline
$ U_1+ e_1$ &$U_1$ & &$U_1+ e_2$ &  $U_1$ &  & $U_1$ \\
$ U_2+e_2$ &$ U_2+ e_1 $ & & $U_2+e_3 $&  $U_2$& & $U_2$\\
$\vdots$ &$\vdots$  & &$\vdots $&$\vdots $&  &$\vdots $\\
$U_{\nNode - \nSysNode -1} + e_{\nNode - \nSysNode-1} $&$U_{\nNode - \nSysNode-1} + e_{\nNode - \nSysNode-2}$ & &$U_{\nNode - \nSysNode-1} + e_{\nNode - \nSysNode} $&$U_{\nNode - \nSysNode-1} $ & &$U_{\nNode - \nSysNode-1} $\\
$U_{\nNode - \nSysNode} + e_{\nNode - \nSysNode}$ &$U_{\nNode - \nSysNode} + e_{\nNode - \nSysNode-1}$ & &$U_{\nNode - \nSysNode}$ & $U_{\nNode - \nSysNode}$ &  &$U_{\nNode - \nSysNode}$ \\
$U_{\nNode - \nSysNode+1}$ & $U_{\nNode - \nSysNode+1}+e_{\nNode - \nSysNode}$ & &$U_{\nNode - \nSysNode+1} $ & $U_{\nNode - \nSysNode+1} $ &  &$U_{\nNode - \nSysNode+1} $ \\
$U_{\nNode - \nSysNode+2} $ &$U_{\nNode - \nSysNode+2}$ & &$U_{\nNode - \nSysNode+2} $ & $U_{\nNode - \nSysNode+2} $ & &$U_{\nNode - \nSysNode+2} $ \\
$\vdots$ &$\vdots$ &  &$\vdots $ & $\vdots $ &  & $\vdots $ \\
$U_{\nSysNode-1}$  &  $U_{\nSysNode-1}$ & &$U_{\nSysNode-1}$ &$U_{\nSysNode-1}$ &  &$U_{\nSysNode-1}$ \\
$U_{\nSysNode}$& $U_{\nSysNode}$ & &$U_{\nSysNode} +e_1$ &  $U_{\nSysNode} $ & &$U_{\nSysNode} $ \\
\hline
\end{tabular}
}
\caption{Queries when user wants $W_1$ and $\nNode-\nSysNode \leq \nSysNode$.}
\vspace{-0.4cm}
\label{tb:query_case1}
\end{table*}
It can be observed that each node receives statistically uniformly random query vectors. Hence, user-privacy is guaranteed.

\noindent{\bf Answers:}
Each node receives $\nSysNode$ query vectors, and for each forms the inner product with the stored data vector, resulting in $\nSysNode$ symbols. Next, they add shared random symbols $S_{i,j}$ to the resulting $\nSysNode$ symbols according to the storage code construction, and send the results to the user. 
Let $X_{i,j} = U_j D_i + S_{i,j}$, where $i \in [1,\nSysNode]$ denotes the index of systematic nodes and $j \in [1,\nSysNode]$ denotes the index of query vectors for each node, there are $\nSysNode^2$ unknowns generated from queries, stored data, and common randomness as follows,
\begin{equation}
{\scriptsize
\begin{bmatrix}
X_{1,1}  & \dots & X_{1,M} \\
\vdots         & \ddots & \vdots                                    \\
X_{M,1}  & \dots & X_{M,M}
\end{bmatrix}
= 
\begin{bmatrix}
U_1 D_1+ S_{1,1}  & \dots & U_M D_1 + S_{1,M} \\
\vdots     & \ddots & \vdots                                    \\
U_1 D_M + S_{M,1} & \dots & U_M D_M + S_{M,M}
\end{bmatrix}}
. \nonumber
\end{equation}
The user receives $\nSysNode$ answers from each node, as shown in Table~\ref{tb:answer_case1}.
\begin{table*}[t]
\centering
\resizebox{\textwidth}{!}{
\begin{tabular}{|c|c|c|c|c|c|c|}
\hline
$A_1^1:$  & $A_2^1: $ & $\dots$ & $A_{\nSysNode}^1: $ & $A_{\nSysNode+1}^1: $ & $\dots$ & $A_{\nNode}^1: $ \\
\hline \hline
$ X_{1,1} + w_ {1,1}^1$ & $X_{2,1}$ &  &$X_{M,1}+ w_{1,M}^2$ &$LC^{M+1}(X_{[1:M],1})$ & &$LC^{N}(X_{[1:M],1})$ \\
$ X_{1,2} + w_{1,1}^2$ &$ X_{2,2}+ w_{1,2}^1 $ & &$X_{M,2}+w_{1,M}^3 $&$LC^{M+1}(X_{[1:M],2})$& & $LC^{N}(X_{[1:M],2})$\\
$\vdots$ & $\vdots$  & & $\vdots $&$\vdots $& &$\vdots $\\ 
$X_{1,\nNode - \nSysNode -1} + w_{1,1}^{\nNode - \nSysNode-1} $ &$X_{2,\nNode - \nSysNode-1} + w_{1,2}^{\nNode - \nSysNode-2}$ & & $X_{M,\nNode - \nSysNode-1} + w_{1,M}^{\nNode - \nSysNode} $ &  $LC^{M+1}(X_{[1:M],\nNode - \nSysNode-1}) $& &  $LC^{N}(X_{[1:M],\nNode - \nSysNode-1}) $\\ 
$X_{1,\nNode - \nSysNode} + w_{1,1}^{\nNode - \nSysNode}$ &$X_{2,\nNode - \nSysNode} + w_{1,2}^{\nNode - \nSysNode-1}$ & &$X_{M,\nNode - \nSysNode}$ & $LC^{M+1}(X_{[1:M],\nNode - \nSysNode}) $ & &$LC^{N}(X_{[1:M],\nNode - \nSysNode}) $ \\  
$X_{1,\nNode - \nSysNode+1}$ &$X_{2,\nNode - \nSysNode+1}+w_{1,2}^{\nNode - \nSysNode}$ & & $X_{M,\nNode - \nSysNode+1} $ &$LC^{M+1}(X_{[1:M],\nNode - \nSysNode+1}) $ &  & $LC^{N}(X_{[1:M],\nNode - \nSysNode+1}) $ \\ 
$X_{1,\nNode - \nSysNode+2} $ & $X_{2,\nNode - \nSysNode+2}$ & &$X_{M,\nNode - \nSysNode+2} $ & $LC^{M+1}(X_{[1:M],\nNode - \nSysNode+2})  $ & &$LC^{N}(X_{[1:M],\nNode - \nSysNode+2})  $ \\ 
$\vdots$ &  $\vdots$ & & $\vdots $ &$\vdots $ & &$\vdots $ \\
$X_{1,\nSysNode-1}$  &   $X_{2,\nSysNode-1}$ & & $X_{M,\nSysNode-1}$ &$LC^{M+1}(X_{[1:M],\nSysNode-1}) $ &  &$LC^{N}(X_{[1:M],\nSysNode-1}) $ \\
$X_{1,\nSysNode}$ & $X_{2,\nSysNode}$& &$X_{M,\nSysNode} +w_{1,M}^1$ & $LC^{M+1}(X_{[1:M],\nSysNode}) $& & $LC^{N}(X_{[1:M],\nSysNode}) $ \\
\hline
\end{tabular}
}
\caption{Answers received by user when $W_1$ is desired and $\nNode-\nSysNode \leq \nSysNode$.}
\vspace{-0.4cm}
\label{tb:answer_case1}
\end{table*}
Note that there are $NM$ unknowns, among which $M^2$ unknowns are the $X_{i,j}$'s and $(N-M)M$ unknowns are symbols of the requested file. It can be observed that there are $NM$ linearly independent equations. Hence, the linear system is solvable. Furthermore, because the user has no information regarding the common randomness, database-privacy is guaranteed.
\begin{table*}[ht]
\centering
\resizebox{\textwidth}{!}{
\begin{tabular}{|c|c|c|c|c|c|c|c|c|}
\hline
$Q_1^1:$  & $Q_2^1: $ & $\dots$ &$Q_{\nSysNode}^1: $ &$Q_{\nSysNode+1}^1 \sim Q_{2 \nSysNode }^1: $ & $Q_{2 \nSysNode +1}^1 \sim Q_{3 \nSysNode}^1 :$ &$\dots$& $Q_{\alpha \nSysNode + 1}^1 \sim Q_{(\alpha+1)\nSysNode}^1:$ & $Q_{(\alpha+1)\nSysNode+1}^1 \sim Q_{\nNode}^1:$ \\
\hline \hline
$ U_1+ e_1$ &$U_1$ & &$U_1+ e_2$ &  $U_1+e_{\beta+1}$ & $U_1+e_{M+\beta+1}$ & & $U_1+e_{(\alpha-1)M+\beta+1}$ & $U_1$  \\
$ U_2+e_2$ &$ U_2+ e_1 $ & & $U_2+e_3 $&  $U_2+e_{\beta+2}$& $U_2+e_{M+\beta+2}$ & &$U_2+e_{(\alpha-1)M+\beta+2}$ &$U_2$ \\
$\vdots$ &$\vdots$  & &$\vdots $&$\vdots $& $\vdots$ & &$\vdots $ & $\vdots$ \\
$U_{\beta -1} + e_{\beta-1} $&$U_{\beta-1} + e_{\beta-2}$ & &$U_{\beta-1} + e_{\beta} $&$U_{\beta-1} +e_{2\beta-1}$ & $U_{\beta-1}+e_{M+2\beta-1}$ & & $U_{\beta-1}+e_{(\alpha-1)M+2\beta-1} $ &$U_{\beta-1} $\\
$U_{\beta} + e_{\beta}$ &$U_{\beta} + e_{\beta-1}$ & &$U_{\beta}$ & $U_{\beta}+e_{2\beta}$ & $U_{\beta}+e_{M+2\beta}$ & &$U_{\beta}+e_{(\alpha-1)M+2\beta}$ & $U_{\beta}$\\
$U_{\beta+1}$ & $U_{\beta+1}+e_{\beta}$ & &$U_{\beta+1} $ & $U_{\beta+1} +e_{2\beta+1}$ & $U_{\beta+1}+e_{M+2\beta+1}$ & &$U_{\beta+1} +e_{(\alpha-1)M+2\beta+1}$  &$U_{\beta+1} $   \\
$U_{\beta+2} $ &$U_{\beta+2}$ & &$U_{\beta+2} $ & $U_{\beta+2} +e_{2\beta+2}$ & $U_{\beta+2}+e_{M+2\beta+2}$ & &$U_{\beta+2} +e_{(\alpha-1)M+2\beta+2}$  & $U_{\beta+2} $ \\
$\vdots$ &$\vdots$ &  &$\vdots $ & $\vdots $ & $\vdots$ & &$\vdots $  & $\vdots$\\
$U_{\nSysNode-1}$  &  $U_{\nSysNode-1}$ & &$U_{\nSysNode-1}$ &$U_{\nSysNode-1}+e_{\beta + M-1}$ & $U_{M-1}+e_{2M+\beta-1}$  & & $U_{\nSysNode-1}+e_{\alpha M +\beta -1}$  &$U_{\nSysNode-1}$ \\
$U_{\nSysNode}$& $U_{\nSysNode}$ & &$U_{\nSysNode} +e_1$ &  $U_{\nSysNode} +e_{\beta+M}$ & $U_M+e_{2M+\beta}$ & &$U_{\nSysNode} +e_{\alpha M +\beta }$  &$U_{\nSysNode} $ \\
\hline
\end{tabular}
}
\caption{Queries when user wants $W_1$ and $\nNode-\nSysNode > \nSysNode$.}
\label{tb:query_case2}
\vspace{-0.5cm}
\end{table*}

\item {\bf Case 2} ($\nNode-\nSysNode > \nSysNode$)

\noindent {\bf Queries:}
Let $\beta = \nNode-\nSysNode \Mod{\nSysNode}$, and $\nNode-\nSysNode = \alpha \nSysNode +\beta$.
The queries are as shown in Table~\ref{tb:query_case2}.
Specifically, for systematic nodes, $\beta$ out of the $\nSysNode$ query vectors retrieve the \emph{first} $\beta$ symbols of the requested file, in a shifted way among all systematic nodes.
The remaining $\alpha \nSysNode$ symbols at each systematic node are retrieved at the parity nodes. Since every $\nSysNode$ symbols need $\nSysNode$ independent linear equations, they are retrieved at $\nSysNode$ parity nodes. 
Similar as in Case 1, user-privacy is guaranteed because nodes receive statistically uniformly random query vectors.

\noindent {\bf Answers:}
The answers are generated in the same way as in Case 1, that is, by forming inner products of the received query vectors and the stored data vectors, and then adding shared random symbols. 
Similarly as in Case 1, the linear system is solvable, hence $W_1$ can be decoded. Besides, database-privacy is guaranteed by the common randomness.

\end{itemize}

\section{Conclusion} 
We study the SPIR problem for coded databases, where a database of $K$ files ($K \geq 2$) is stored at $N$ nodes based on an $(N,M)$-MDS storage code. A user wants to retrieve one file without revealing the identity of the requested file to the nodes. At the same time, the user shall obtain no more information regarding the database other than the requested file. 
We derive the SPIR capacity for coded databases to be $1-\frac{M}{N}$, where $\frac{M}{N}$ is the rate of the MDS storage code. To achieve this capacity or any positive rate for SPIR, the storage nodes need to share common randomness that is unavailable to the user and independent of the database, with amount at least $\frac{M}{N-M}$ times the file size.

\bibliographystyle{IEEEtran}
\bibliography{IEEEabrv,SPIR}

\begin{thebibliography}{10}
\providecommand{\url}[1]{#1}
\csname url@samestyle\endcsname
\providecommand{\newblock}{\relax}
\providecommand{\bibinfo}[2]{#2}
\providecommand{\BIBentrySTDinterwordspacing}{\spaceskip=0pt\relax}
\providecommand{\BIBentryALTinterwordstretchfactor}{4}
\providecommand{\BIBentryALTinterwordspacing}{\spaceskip=\fontdimen2\font plus
\BIBentryALTinterwordstretchfactor\fontdimen3\font minus
  \fontdimen4\font\relax}
\providecommand{\BIBforeignlanguage}[2]{{%
\expandafter\ifx\csname l@#1\endcsname\relax
\typeout{** WARNING: IEEEtran.bst: No hyphenation pattern has been}%
\typeout{** loaded for the language `#1'. Using the pattern for}%
\typeout{** the default language instead.}%
\else
\language=\csname l@#1\endcsname
\fi
#2}}
\providecommand{\BIBdecl}{\relax}
\BIBdecl

\bibitem{chor1998private}
B.~Chor, E.~Kushilevitz, O.~Goldreich, and M.~Sudan, ``Private information
  retrieval,'' \emph{Journal of the ACM (JACM)}, vol.~45, no.~6, pp. 965--981,
  1998.

\bibitem{gertner1998protecting}
Y.~Gertner, Y.~Ishai, E.~Kushilevitz, and T.~Malkin, ``Protecting data privacy
  in private information retrieval schemes,'' in \emph{Proceedings of the
  thirtieth annual ACM symposium on Theory of computing}, 1998, pp. 151--160.

\bibitem{gasarch2004survey}
W.~Gasarch, ``A survey on private information retrieval,'' in \emph{Bulletin of
  the EATCS}.\hskip 1em plus 0.5em minus 0.4em\relax Citeseer, 2004.

\bibitem{sun2016capacity}
H.~Sun and S.~A. Jafar, ``The capacity of private information retrieval,''
  \emph{arXiv preprint arXiv:1602.09134}, 2016.

\bibitem{sun2016colluding}
------, ``The capacity of robust private information retrieval with colluding
  databases,'' \emph{arXiv preprint arXiv:1605.00635}, 2016.

\bibitem{sun2016SPIR}
------, ``The capacity of symmetric private information retrieval,''
  \emph{arXiv preprint arXiv:1606.08828}, 2016.

\bibitem{dimakis2010network}
A.~G. Dimakis, P.~B. Godfrey, Y.~Wu, M.~J. Wainwright, and K.~Ramchandran,
  ``Network coding for distributed storage systems,'' \emph{IEEE Transactions
  on Information Theory}, vol.~56, no.~9, pp. 4539--4551, 2010.

\bibitem{shah2014one}
N.~B. Shah, K.~Rashmi, and K.~Ramchandran, ``One extra bit of download ensures
  perfectly private information retrieval,'' in \emph{Proceedings of IEEE
  International Symposium on Information Theory (ISIT)}, 2014, pp. 856--860.

\bibitem{fazeli2015pir}
A.~Fazeli, A.~Vardy, and E.~Yaakobi, ``{PIR} with low storage overhead: coding
  instead of replication,'' \emph{arXiv preprint arXiv:1505.06241}, 2015.

\bibitem{chan2015private}
T.~H. Chan, S.-W. Ho, and H.~Yamamoto, ``Private information retrieval for
  coded storage,'' in \emph{Proceedings of IEEE International Symposium on
  Information Theory (ISIT)}, 2015, pp. 2842--2846.

\bibitem{tajeddine2016private}
R.~Tajeddine and S.~E. Rouayheb, ``Private information retrieval from {MDS}
  coded data in distributed storage systems,'' in \emph{Proceedings of IEEE
  International Symposium on Information Theory (ISIT)}, 2016.

\bibitem{banawan2016capacity}
K.~Banawan and S.~Ulukus, ``The capacity of private information retrieval from
  coded databases,'' \emph{arXiv preprint arXiv:1609.08138}, 2016.

\end{thebibliography}

\end{document}